\documentclass{aa}
\usepackage{natbib}
\usepackage{amsmath,amssymb}
\usepackage{mathabx,mathrsfs}
\usepackage{natbib,twoopt}
\usepackage{graphicx}
\usepackage{hyperref}
\usepackage{svg}
\usepackage{txfonts}
\usepackage{atbegshi}
\usepackage{tablefootnote}
\usepackage[para,online,flushleft]{threeparttable}
\usepackage{ulem} 
\bibpunct{(}{)}{;}{a}{}{,}

\begin{document}

\newcommand{\kms}{km\,s$^{-1}$}
\newcommand{\teff}{$T_\mathrm{eff}$}
\newcommand{\logg}{$\log g$}
\newcommand{\feh}{[Fe/H]}
\newcommand{\mh}{[M/H]}
\newcommand{\afe}{[$\alpha$/Fe]}
\newcommand{\am}{[$\alpha$/M]}
\newcommand{\cm}{[C/M]}
\newcommand{\vmicro}{$v_{\rm micro}$}
\newcommand{\mum}{$\mu$m}
\newcommand{\ebv}{$E(B-V)$}
\newcommand{\av}{$A_V$}

\title{The Gothard Observatory Synthetic Stellar Photometry Database}

\author{
J{\'o}zsef Kov{\'a}cs\inst{1,2,3},
Szabolcs~M{\'e}sz{\'a}ros\inst{1,2,4},
Be{\'a}ta~Harmati\inst{1,2},
Borb{\'a}la Cseh\inst{2,4},
Viola~Heged{\H{u}}s\inst{1,2},
Gyula~M.~Szab{\'o}\inst{1,2,3},
L{\'a}szl{\'o} Szigeti\inst{1,2},
Blanka Vil{\'a}gos\inst{1,2},
Aliz~Derekas\inst{1,2,5}
}

\institute{
ELTE E\"otv\"os Lor\'and University, Gothard Astrophysical Observatory, Szent Imre H. u. 112., Szombathely H-9700, Hungary
\and
MTA--ELTE Lend{\"u}let "Momentum" Milky Way Research Group, Szent Imre H. u. 112., Szombathely H-9700, Hungary
\and
HUN--REN--ELTE Exoplanet Research Group, Szent Imre H. u. 112., Szombathely H-9700, Hungary
\and
HUN-REN CSFK, Konkoly Observatory, Konkoly Thege Mikl\'os \'ut 15-17, Budapest, H-1121, Hungary
\and
HUN--REN Stellar Astrophysics Research Group, Szegedi \'ut, Kt. 766, Baja, H-6500, Hungary
}

\date{Submitted December 17, 2025}

\abstract
{In order to determine stellar luminosities and radii, it is necessary to know the total bolometric fluxes emitted by the stars, or equivalently the bolometric corrections (BCs) as accurately as possible.}
{The aim of this paper is to present and describe a new database of synthetic stellar magnitudes and bolometric corrections for 752 filters from 78 ground- and space-based instruments calculated using the most recent version of the BOSZ synthetic stellar spectral library.}
{From the entire grid of the BOSZ theoretical spectra, our synthetic magnitudes in the Vega magnitude system were determined using the corresponding routines of the Python package {\tt species}.}
{The database spans effective temperatures from 2800 to 16000~K, \logg\ from $-0.5$ to 5.5, metallicities from $-2.5$ to 0.75, \am\ from $-0.25$ to 0.5, \cm\ from $-0.75$ to 0.5, and reddening up to $A_V$ = 3.1 mag. Using high-resolution (R = 50000) synthetic spectra allowed us to precisely track the effect of abundances on the bolometric corrections and luminosity of stars.}
{By applying the new bolometric corrections (BCs) to 192\,000 APOGEE stars we calculated luminosities, and also demonstrated that neglecting carbon can introduce up to $\pm$0.2\% errors in luminosity. The new Gothard Observatory Synthetic Stellar Photometry Database may enable more accurate fundamental parameter determinations for large stellar samples using a vast amount of past, present, and upcoming surveys, such as Gaia, LSST, and the Roman Space Telescope.}


\titlerunning{New Gaia DR3 Bolometric Corrections}
\authorrunning{Kov{\'a}cs et al. 2025}
\maketitle

\section{Introduction}
\label{sec:introduction}

Synthetic stellar photometry is an essential method/tool for relating theoretical stellar parameters, namely effective temperature (\teff), surface gravity (\logg), metallicity (\mh) and often alpha element abundances (\am), to observable quantities, e.g. magnitudes and colors in various photometric systems. In order to determine the stellar luminosities and radii, it is necessary to know the total bolometric fluxes emitted by the stars, or equivalently the bolometric corrections (BCs). The recent versions of BCs are usually based on theoretical spectral libraries calculated using stellar atmosphere models and atomic and molecular line lists.

\cite{2014MNRAS.444..392C} gave an overview of the synthetic photometry, and also included the case of 2MASS, SDSS, HST, and Johnson-Cousins filters, while \cite{2018MNRAS.475.5023C} discussed in detail the application of synthetic photometry using MARCS stellar atmosphere models based on the solar chemical composition by \cite{2007SSRv..130..105G} to determine synthetic colors, bolometric corrections and reddening coefficients for the Hipparcos/Tycho, Pan-STARRS1 \citep{chambers2019panstarrs1surveys}, SkyMapper \citep{Keller_Schmidt_Bessell_Conroy_Francis_Granlund_Kowald_Oates_Martin-Jones_Preston_et_al._2007} and JWST photometric systems containing 85 filters in total. Using absolute spectrophotometry from the CALSPEC library \citep{2014PASP..126..711B}, they showed that bolometric fluxes can be recovered to about 2 percent from bolometric corrections in a single band.

Expanding their previous work, a similar study was carried out by \cite{2018MNRAS.479L.102C} for the Gaia \citep{2016A&A...595A...1G} $G_\mathrm{BP}$, $G$ and $G_\mathrm{RP}$ bands based on the Gaia DR2 \citep{2018A&A...616A...1G} and also the entire grid of the MARCS models. They investigated the effects of slightly different transmission curves (,,processed'' and ,,revised'') and found that the differences are typically of a few millimagnitudes only, but found a magnitude-dependent offset in Gaia $G$ magnitudes. They concluded that the magnitudes of $G$ and $G_\mathrm{RP}$ are typically better than $G_\mathrm{BP}$ in recovering bolometric fluxes.\footnote{An updated version of their BCs database for Gaia DR3 is available on GitHub page of L. Casagrande, see \url{https://github.com/casaluca/bolometric-corrections}}

\cite{2019A&A...632A.105C} presented the YBC database of stellar bolometric corrections, in which they homogenized widely used theoretical stellar spectral libraries and provided BCs for many (at least 70) popular photometric systems, including Gaia (DR2) filters. They computed BC tables both with and without extinction, therefore, the YBC database provides a more realistic way to fit isochrones with spectral-type dependent extinction. The effect of taking reddening into account in the bolometric correction is very clearly seen in their Fig.~4, which shows that at $A_V = 0.5$~mag there is already $\sim$0.2~mag difference in the Gaia $G$ band, $\sim$0.1~mag in the Gaia $G_\mathrm{BP} - G_\mathrm{RP}$ color, and $\sim$1~mag in the HST/WFC3 F218W filter. Therefore, they suggested using extinction coefficients dependent on the spectral type for Gaia filters and UV filters whenever $A_V \gtrsim 0.5$~mag.

\begin{table}
\caption{List of the facilities, instruments and number of the filters involved in the bolometric correction and synthetic magnitude calculations following the name convention of the SVO website.}
\label{fac_inst_filt_num}
\begin{center}
\begin{tabular}{p{1.1cm}p{1.8cm}rp{1.1cm}p{1.8cm}r}
\hline
{\bf Facility} & {\bf Instrument} & {\bf N} &
{\bf Facility} & {\bf Instrument} &  {\bf N} \\
\hline
2MASS       & 2MASS             &  3 & KPNO         & DESI          &  3 \\
AKARI       & IRC               &  9 &              & TIFKAM        &  4 \\
APO         & Broad             &  5 & LaSilla      & WFI           & 23 \\
            & SDSS              &  5 & LBT          & LBCB          &  7 \\
CAHA        & Omega2000         & 15 &              & LBCR          & 13 \\
CASTOR      & CASTOR0           &  2 & LCO          & OGLE-IV       &  2 \\
CFHT        & MegaCam           & 12 &              & WFCCD         &  4 \\
            & Wircam            & 11 & LSST         & LSST          &  6 \\
Euclid      & NISP              &  3 & NOT          & NOTcam        & 16 \\
            & VIS               &  1 & Palomar      & Arp1961       &  4 \\
GAIA        & GAIA2r            &  3 &              & GSC2          &  3 \\
            & GAIA3             &  4 &              & POSS          &  2 \\
GALEX       & GALEX             &  1 &              & Wevers1986    &  3 \\
Gemini      & DSSI              &  4 &              & ZTF           &  3 \\
            & GSAOI             & 20 & PAN-STARRS   & PS1           &  7 \\
            & Michelle          &  8 & Paranal      & ISAAC         &  9 \\
Generic     & Bessell           &  5 &              & OmegaCAM      & 24 \\
            & Bessell\_JHKLM    &  6 &		        & VISTA         &  5 \\
            & Cousins           &  2 & Roman        & WFI           & 10 \\
            & Johnson           &  7 & SkyMapper    & SkyMapper     &  6 \\
            & Stromgren         &  4 & SLOAN        & SDSS          &  5 \\
GTC         & CanariCam         & 18 & Spitzer      & IRAC          &  4 \\
            & OSIRIS            & 30 &              & IRS           &  2 \\
Hipparcos   & Hipparcos         &  3 &              & MIPS          &  1 \\
HST         & ACS\_HRC          & 18 & Subaru       & CIAO          & 17 \\
            & ACS\_WFC          & 13 &              & FOCAS         &  5 \\
            & NICMOS1           & 11 &              & HSC           &  5 \\
            & NICMOS2           & 17 &              & IRCS          &  6 \\
            & NICMOS3           & 12 &              & MOIRCS        &  5 \\
            & WFC3\_IR          & 13 &              & OHS           &  9 \\
            & WFC3\_UVIS1       & 28 &              & Suprime       & 26 \\
            & WFC3\_UVIS2       & 28 & TYCHO        & TYCHO         &  6 \\
            & WFPC2-PC          & 24 & UKIRT        & UKIDSS        &  5 \\
            & WFPC2-WF          & 26 &              & WFCAM         &  5 \\
IRAS        & IRAS              &  2 & WHT          & INGRID        & 13 \\
JWST        & MIRI              &  9 &              & LRIS          & 13 \\
            & NIRCam            & 29 &              & PAUCam        &  6 \\
            & NIRISS            & 12 &              & PFIP          &  2 \\
Keck        & ESI               &  3 & WISE         & WISE          &  4 \\
            & LRIS              & 11 &              &               &    \\
            & LWS               & 11 &              &               &    \\
            & NIRC2             & 11 &              &               &    \\
\end{tabular}
\end{center}
\end{table}

The only database of BC corrections to Gaia $G$ magnitudes for Gaia DR3 \citep{2023A&A...674A...1G} was published by \cite{2023A&A...674A..26C}. They provided a Python-implemented bolometric correction function which takes as input the effective temperature \teff, surface gravity \logg, iron abundance \feh, and alpha-enhancement \afe, and returns the model value $\mathrm{BC}_G$ for that combination of parameters. The $\mathrm{BC}_G$ values for interpolation were derived from the synthetic stellar spectra based on a grid of MARCS models \citep{2008A&A...486..951G} supplemented with intermediate temperature stars \citep{2005MSAIS...7...99S} as a function of \teff, \logg, \feh, and \afe. They assumed \afe\ = 0.0 when calculating the correction for all stars because \afe\ was only estimated for a small fraction of the sources. Their adopted value for the bolometric correction for the Sun is $\mathrm{BC}_{G,\odot} = +0.08$~mag, where $M_{\mathrm{bol},\odot} = 4.74$ yields an absolute magnitude of the Sun $M_{G,\odot} = 4.66$~mag. For extinction calculations they used the wavelength-dependent extinction law by \citet{1999PASP..111...63F}.

In this work, we use the BOSZ synthetic spectral library originally developed to flux calibrate the James Webb Space Telescope \citep{2024A&A...688A.197M} to compute new magnitudes and bolometric corrections for 752 filters from 78 instruments used in 35 facilities.\footnote{The database can be downloaded in part or in its entirety from \url{https://gosspd.gothard.hu/data}.} We calculate new luminosity for about 192\,000 APOGEE stars for the Gaia DR3 $G$ filter. The paper is structured as follows. We summarize the calculation methods, including the scope of the database in Section~\ref{sec:calculation_methods}, followed by the validation aspects of our database in Section~\ref{sec:validating_the_database}. Next we compare our new Gaia $G$ BCs with the set of official Gaia DR3 $G$ values in Section~\ref{sec:comparision_with_gaia_dr3}, while in Section~\ref{sec:apogee_luminosities} we present our new luminosity values for a large sample of APOGEE stars. Finally, we provide a brief overview of our results in Section~\ref{sec:overview}.

\section{Calculation methods}
\label{sec:calculation_methods}

Synthetic photometry is the process of deriving brightness and color values by convolving fluxes from stellar atmospheric models with filter functions of standard photometric passbands. The response function of a standard photometric system is usually obtained by multiplying the function describing the reflectivity of the telescope mirror by the transfer function of the filters and camera optical system and the quantum efficiency of the detector as a function of the wavelength. For ground-based observations, this must be multiplied by the transfer function of the Earth's atmosphere for an air mass of at least 1.0. Atmospheric correction is usually performed in the broad UV bands (like the $U$ band), where extinction is large and varies significantly within the band, also in the red edge of the optical range, and in the IR bands, where absorption by molecules (e.g. H$_2$O and O$_2$) is significant. Previous summaries on synthetic photometry methods are given in \cite{1976MmRAS..81...25C}, \cite{1986HiA.....7..799B}, \cite{1996BaltA...5..459S}, and \cite{1996AJ....112.2274C}, while more recently a pedagogical introduction to the main concepts is given in \cite{2014MNRAS.444..392C}.

\subsection{From fluxes to magnitudes}

In its most common and simplest form, the synthetic magnitude is proportional to the logarithm of
\begin{equation}
\frac{\int_{x_a}^{x_b} f_x T_\zeta \, dx}{\int_{x_a}^{x_b} T_\zeta \, dx},
\end{equation}
where $f_x$ is the flux at $x$, $T_\zeta$ is the system response function, and the integration over $x$ is carried out in the wavelength ($x = \lambda$) or frequency ($x = \nu$) space between the limits $x_a$ and $x_b$. The system response function $T_\zeta$ represents the total throughput, which is affected by everything between the top of the Earth's atmosphere and the final detection of the photon; hence the proper characterization of $T_\zeta$ is a crucial and non-trivial part of most photometric systems.

Although we work with fluxes as physical quantities, in astronomy we use the magnitude scale created by Hipparcos and later formalized by Pogson to characterize the brightness of stars. According to the Pogson formula, the brightness of a star through the filter $i$ is
\begin{equation}
m_i = - 2.5 \log \frac{\int_{x_a}^{x_b} f_{x} T_\zeta \, dx}{\int_{x_a}^{x_b} f_{x}^{0} T_\zeta \, dx} + m_{i,0},
\end{equation}
where $f_{x}^{0}$ and $m_{i,0}$ are the flux and magnitude values of a standard star (or reference spectrum), respectively, and $\log$ is the logarithm with base 10.

\begin{figure*}
\centering
\includegraphics[width=7.1in,angle=0]{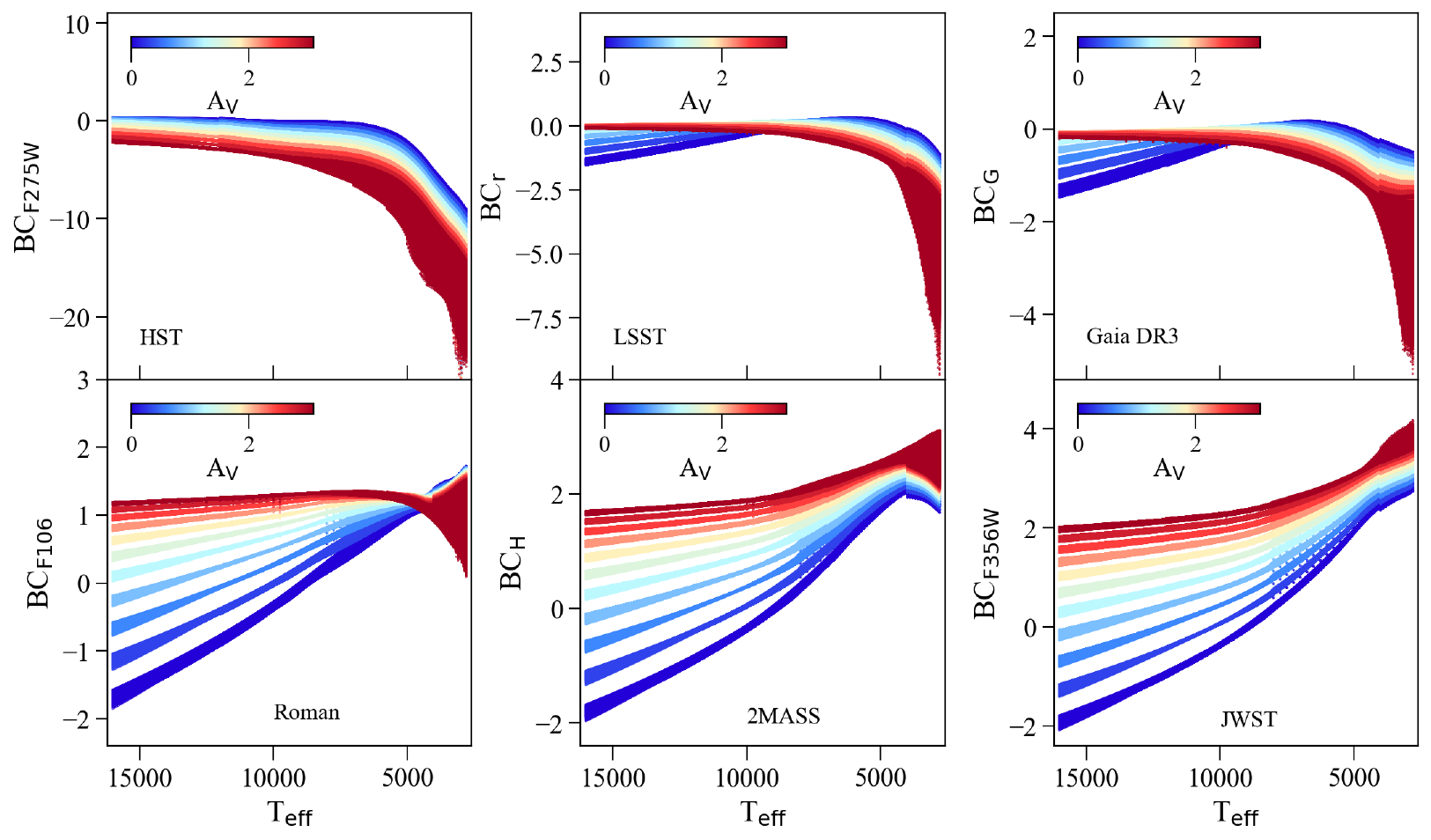}
\caption{Examples of how bolometric corrections in selected HST, LSST, Gaia, Roman, 2MASS and JWST filters depend on effective temperature. Filters are shown in the order of increasing central wavelength. The bolometric corrections are color coded by reddening.}
\label{teff_bcs}
\end{figure*}

Since the flux is determined after the source signal has been recorded by the detector, a distinction must be made between photon-counting and energy-integrating detectors. The CCD cameras used today almost exclusively count the photons arriving at the detector surface per unit time; thus, in the formula defining the magnitude, it is useful to divide the flux $f_\lambda$ by the photon energy $h\nu = hc/\lambda$. Without going into details, this means that, in the formulas, we have to replace $T_\zeta$ by $\lambda T_\zeta$. (The $hc$ constant disappears due to normalization over $T_\zeta$.) So, if we use the wavelength, the definition of the brightness of a star is
\begin{equation}
m_i = - 2.5 \log \frac{\int_{\lambda_1}^{\lambda_2} \lambda f_\lambda T_{\lambda,i} \, d\lambda}{\int_{\lambda_1}^{\lambda_2} \lambda f_\lambda^0 T_{\lambda,i} \, d\lambda} + m_{i,0}.
\end{equation}
Depending on the flux of the reference star or the spectra $f_\lambda^0$ ($f_\nu^0$), the most widely used magnitude systems are the following.

\begin{itemize}

\item {\it Vega magnitude system}.
The spectrum of Vega ($\alpha$~Lyr) is used as a reference spectrum. The reference magnitudes are set so that Vega has a magnitude equal to or slightly different from zero. The latest Vega spectrum is available in the CALSPEC\footnote{\url{https://ssb.stsci.edu/cdbs/calspec/alpha\_lyr\_stis\_011.fits}} database \citep{2014PASP..126..711B}.

\item {\it AB magnitude system} \citep{1974ApJS...27...21O}.
The reference spectrum of the AB magnitude system has a constant value of $f_\nu^0 = 3.63 \times 10^{-20} \, \mathrm{erg} \, \mathrm{s}^{-1} \, \mathrm{cm}^{-2} \, \mathrm{Hz}^{-1}$. The reference magnitudes are therefore set to $-2.5 \log f_\nu^0 = -48.6$~mag.

\item {\it ST magnitude system} \citep{2018ascl.soft11001S}.
The reference spectrum of the ST magnitude system has a constant value of $f_\lambda^0 = 3.63 \times 10^{-9} \, \mathrm{erg} \, \mathrm{s}^{-1} \, \mathrm{cm}^{-2} \, \mathring{\mathrm{A}}^{-1}$. The reference magnitudes are therefore set to $-2.5 \log f_\lambda^0 = -21.1$~mag.

\end{itemize}

It is worth noting that in the AB system, where frequency is used instead of wavelength, $f_{\lambda}^{(0)} \, d\lambda$ should not simply be replaced with $f_{\nu}^{(0)} \, d\nu$, but with $-f_{\nu}^{(0)} \frac{c}{\nu^2} \, d\nu$.

\subsection{Bolometric corrections}

Synthetic spectral libraries provide the stellar flux $F_\lambda$ (usually in $\mathrm{erg} \, \mathrm{s}^{-1} \, \mathrm{cm}^{-2} \, \mathring{\mathrm{A}}^{-1}$) on a surface element for a given set of stellar parameters. The flux $F_\lambda$ is related to the effective temperature \teff\ of the star through the form
\begin{equation}
F_\mathrm{bol} = \int_0^\infty F_\lambda \, d\lambda = \sigma T^4_\mathrm{eff},
\end{equation}
where $\sigma$ is the Boltzmann constant. If the distance to the star from Earth is $d$, then the flux we can measure is
\begin{equation}
f_{\lambda,d} = \left ( \frac{R}{d} \right )^2 F_\lambda 10^{-0.4 A_\lambda},
\end{equation}
where $A_\lambda$ is the extinction between the star and the observer. If distance $d$ is chosen to be 10~pc, the absolute magnitude $M_i$ for a photon-counting photometric system is
\begin{eqnarray}
M_i & = & - 2.5 \log \frac{\int_{\lambda_1}^{\lambda_2} \lambda f_{\lambda,10\,\mathrm{pc}} T_{\lambda,i} \, d\lambda}{\int_{\lambda_1}^{\lambda_2} \lambda f_\lambda^0 T_{\lambda,i} \, d\lambda} + m_{i,0} \nonumber \\
& = & -2.5 \log
\left [
\left ( \frac{R}{10\,\mathrm{pc}} \right )^2
\frac{\int_{\lambda_1}^{\lambda_2} \lambda F_\lambda 10^{-0.4 A_\lambda} T_{\lambda,i} \, d\lambda}{\int_{\lambda_1}^{\lambda_2} \lambda f_\lambda^0 T_{\lambda,i} \, d\lambda}
\right ] + m_{i,0}.
\end{eqnarray}
The definition of bolometric magnitude $M_\mathrm{bol}$ is
\begin{eqnarray}
M_\mathrm{bol} & = & M_{\mathrm{bol},\odot} - 2.5 \log (L / L_\odot) \nonumber \\
& = & M_{\mathrm{bol},\odot} -2.5 \log (4\pi R^2 F_\mathrm{bol} / L_\odot).
\end{eqnarray}
According to the IAU 2015 resolution \citep{2015arXiv151006262M}, the absolute bolometric magnitude for the nominal solar luminosity of $3.828 \times 10^{26} \, \mathrm{W}$ is $M_{\mathrm{bol},\odot} = 4.74\,\mathrm{mag}$.

Given an absolute magnitude $M_i$ in the filter band $i$ for a star of absolute bolometric magnitude $M_\mathrm{bol}$, the bolometric correction $\mathrm{BC}_i$ is
\begin{equation}
\mathrm{BC}_i = M_\mathrm{bol} - M_i = m_\mathrm{bol} - m_i.
\end{equation}
It can be shown that the bolometric flux of a star having an observed magnitude $m_i$ and the bolometric correction $\mathrm{BC}_i$ is
\begin{equation}
f_\mathrm{bol} =
\frac{\pi L_\odot}{(1.296 \times 10^7 \, \mathrm{au})^2} \,
10^{-0.4 (\mathrm{BC}_i - M_{\mathrm{bol},\odot} + m_i)}.
\label{eq:Cas_f_bol}
\end{equation}
This equation is equivalent to that of \cite{2018MNRAS.475.5023C}. The solar luminosity and the astronomical unit should be given in $\mathrm{erg}\,\mathrm{s}^{-1}$ and cm, respectively. The adopted value for au from IAU Resolution B2 is $\mathrm{au} = 1.495978707 \times 10^{13} \, \mathrm{cm}$.

The new BOSZ synthetic spectral library provides surface brightness values ($H$) in $\mathrm{erg} \, \mathrm{s}^{-1} \, \mathrm{cm}^{-2} \, \mathring{\mathrm{A}}^{-1}$, which can be converted to the flux in the proper format given in $\mathrm{W}\,\mathrm{m}^{-2}$ using the equation
\begin{equation}
f = 10 \times 4 \pi H \left ( \frac{R_\odot}{1\,\mathrm{au}} \right )^2
\end{equation}
Once we have this flux from the synthetic spectrum, we can calculate the bolometric correction using Eq. (\ref{eq:Cas_f_bol}) as
\begin{equation}
\mathrm{BC}_i = -2.5 \log_{10} \frac{f_\mathrm{bol}}{3.1993 \times 10^{-11}} - m_i + 4.74,
\label{eq:bc}
\end{equation}
where $3.1993 \times 10^{-11}$ is the numerical value of the expression containing $\pi$, $L_\odot$, and $\mathrm{au}$. All BCs presented in this paper have been calculated with equation (\ref{eq:bc}). Now, taking into account interstellar reddening, the absolute bolometric magnitude can be determined as
\begin{equation}
M_\mathrm{bol} = m_i + \mathrm{BC}_i + 5 - 5 \log d - A_i,
\end{equation}
where $d$ is the distance to the star in parsecs and $A_i$ is the interstellar extinction. Given the absolute bolometric magnitude, there is a straightforward way to calculate the luminosity of a star with the following equation.
\begin{equation}
\frac{L}{L_\odot} = 10^{(M_{\mathrm{bol},\odot} - M_\mathrm{bol}) / 2.5}
\label{L_per_Lsun}
\end{equation}
Knowing the luminosity, the radius of the star can be determined from the Stefan-Boltzmann law as:
\begin{equation}
\frac{R}{R_\odot} = \sqrt{\frac{L}{L_\odot}} \, \frac{T_{\mathrm{eff},\odot}^2}{T_\mathrm{eff}^2}
\label{R_per_Rsun}
\end{equation}

\subsection{Parameters of the new BC database}

\begin{scriptsize}
\begin{table}
\caption{List of the grid parameters used for magnitude and BC calculations.}
\label{grid_parameters_for_BC_calc}
\begin{center}
\begin{tabular}{cccc}
\hline
grid & lower & upper & step \\
parameter & boundary & boundary & size \\
\hline
\teff\ (K) & 2800 & 16\,000 & 50 \\
\logg & $-0.50$ & $5.50$ & $0.50$ \\
\mh & $-2.50$ & $0.75$ & $0.25$ \\
\afe & $-0.25$ & $0.50$ & $0.25$ \\
\cm & $-0.75$ & $0.50$ & $0.25$ \\
$A_V$ & $0.00$ & $3.10$ & $0.31$ \\
\hline
\end{tabular}
\end{center}
\end{table}
\end{scriptsize}

\begin{scriptsize}
\begin{table}
\caption{Synthetic magnitudes and bolometric corrections of the Sun following the name convention of the SVO website. This table in its entirety is available in machine readable form from \url{https://gosspd.gothard.hu/data}.}
\begin{center}
\begin{tabular}{p{2.8cm}cp{0.0cm}cp{0.0cm}}
\hline
Filter & Absolute & & Bolometric \\
ID & magnitude & & correction \\
\hline
2MASS/2MASS.H  & 3.3655 & & 1.3931 \\
2MASS/2MASS.J  & 3.6792 & & 1.0504 \\
2MASS/2MASS.Ks & 3.3024 & & 1.4272 \\
AKARI/IRC.L15  & 3.2689 & & 1.4607 \\
AKARI/IRC.L18W & 3.2670 & & 1.4626 \\
\hline
\end{tabular}
\label{solar_bcs_and_mags}
\end{center}
\end{table}
\end{scriptsize}

Our new bolometric correction calculations are based on the BOSZ synthetic spectral library \citep{2017AJ....153..234B, 2024A&A...688A.197M}, which was developed to flux calibrate the James Webb Space Telescope. \citet{2024A&A...688A.197M} gives a detailed description of the updated version, which is used for all calculations presented in this paper.

The new grid was calculated with {\tt synspec} \citep{hubeny2021tlustysynspecuserssguide} using the LTE approximation and covers metallicities \mh\ from $-2.5$ to 0.75~dex, \afe\ from $-0.25$ to 0.5~dex and \cm\ from $-0.75$ to 0.5~dex, providing synthetic spectra for 336 unique compositions. Calculations for stars between 2800 and 8000 K use MARCS model atmospheres \citep{2008A&A...486..951G}, and ATLAS9 \citep{1979ApJS...40....1K, 2012AJ....144..120M} is used between 7500 and 16,000~K. Examples of bolometric corrections in selected filters are shown in Fig.~\ref{teff_bcs}.

The new BOSZ grid includes 628,620 synthetic spectra from 50 nm to 32 µm with models for 495 \teff\ -- \logg\ parameter pairs per composition and per microturbulent velocity. Each spectrum is available in eight different resolutions that span a range of R = 500 to 50,000 as well as the original resolution of the synthesis. The microturbulent velocities are 0, 1, 2, and 4 km\,s$^{-1}$. In this paper, the R = 50,000 spectra with 2 km\,s$^{-1}$ microturbulent velocity were selected for all model spectra. In order to carefully test the effect of carbon and $\alpha$-abundances on the synthetic magnitudes and BCs, we derived these values using spectra at various resolutions (5000, 10,000, 20,000, 50,000) in the BOSZ database. It was concluded that while the magnitude differences were small across the various resolutions, the R = 50,000 version was the most ideal to precisely track the effect of $\alpha$ and carbon abundances on the stellar spectra.

\begin{figure*}
\centering
\includegraphics[width=6.2in,angle=0]{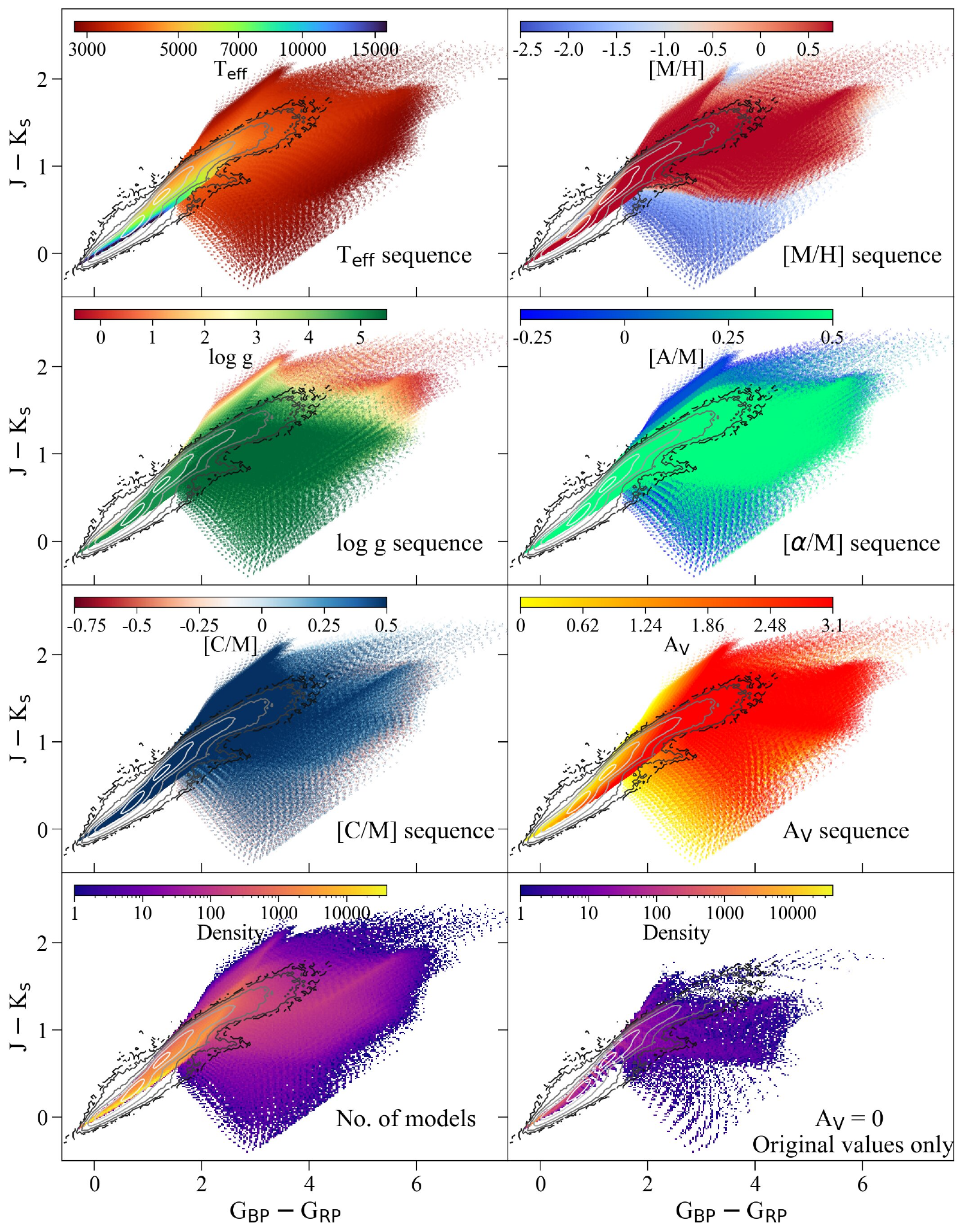}
\caption{The J$-$K$_{\rm s}$ vs. G$_{\rm BP}-$G$_{\rm RP}$ color-color diagram. The contoured area is stars observed by both Gaia and 2MASS, see Section~\ref{validate} for more detail. Each panel is color-coded by a different parameter indicated in the panels themselves.}
\label{color2}
\end{figure*}

In the 2024 version of the BOSZ grid, significant updates were made to the molecular line list compared to the old version by including line lists from the ExoMol project \citep{2012MNRAS.425...21T,2016JMoSp.327...73T}. The new BOSZ grid includes 23 molecules (AlH, AlO, C$_2$, CaH, CaO, CH, CN, CO, CrH, FeH, H$_2$, H$_2$O, MgH, MgO, NaH, NH, OH, OH+, SiH, SiO, TiH, TiO, and VO). The line lists of several molecules in the previous version have been updated. The newer line lists contain more lines, mainly in the infrared region, allowing more accurate spectral modeling. In principle, more accurate synthetic magnitudes and bolometric corrections can be calculated from higher quality synthetic spectra.

\subsection{Fixes made in the 2024 version of the BOSZ grid}

During the calculation of the Gothard Observatory Synthetic Stellar Photometry Database, the team discovered some issues with the 2024 version of the BOSZ grid, which have been fixed and the BOSZ website\footnote{https://archive.stsci.edu/hlsp/bosz} has also been updated. One of the issues was relatively minor; an 8-level H atom was used originally, which resulted in a lack of higher-level H lines above 5.8 microns. For the fix, we used a 40-level H atom, which allowed the correct calculation of all hydrogen lines up to 32 microns. The second issue was related to OH+. Unfortunately, the OH partition table was used for OH+ in the 2024 version, resulting in stronger than expected molecular absorption lines between 300 and 450 nm, and 1 and 8 microns for models with temperatures below 6500K. This issue has also been fixed in the code providing, now the correct OH+ line strength. All bolometric corrections and synthetic magnitudes presented in this paper are based on fixed 2024 BOSZ spectra.

\subsection{Interstellar reddening}

Interstellar reddening was also taken into account when determining bolometric corrections. We used the {\tt ccm89} reddening function of the Python package {\tt extinction}\footnote{\url{http://github.com/kbarbary/extinction}}. This function uses the reddening law determined by \citet{1989ApJ...345..245C}. After carrying out several numerical tests, we decided not to redden the bolometric corrections but rather to redden the synthetic spectra themselves before determining the bolometric corrections. Reddening the spectra was achieved by varying $A_V$ between 0.0 and 3.1 in 11 steps with a step size of 0.31 in equation $R_V = A_V / E(B-V)$, where $R_V$ was set to 3.1. This resulted in having 0.1 step size in $E(B-V)$ between $E(B-V)$ = 0 and 1.

\begin{figure*}
\centering
\includegraphics[width=7.2in,angle=0]{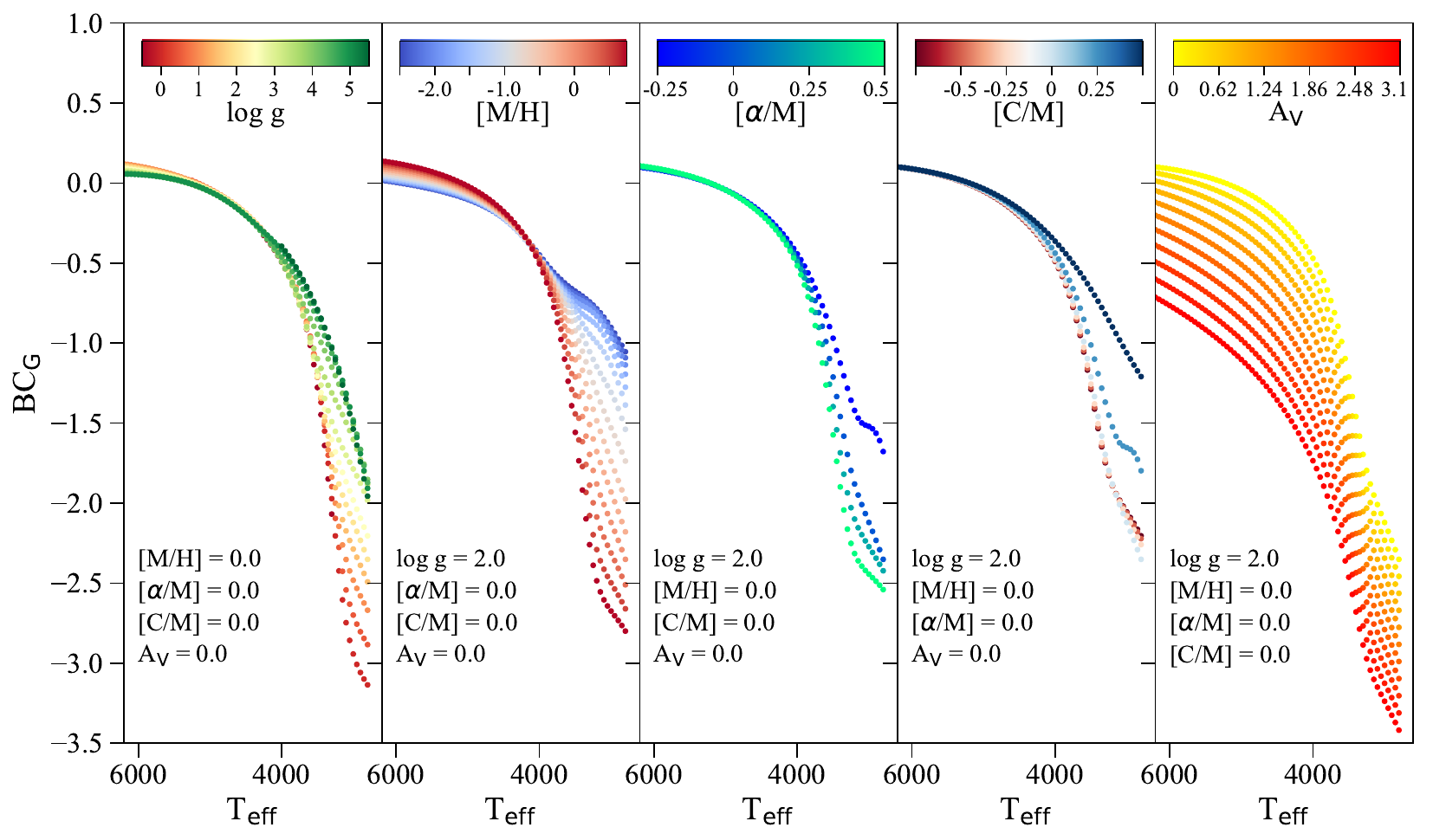}
\caption{Effect of various parameters on Gaia DR3 $\mathrm{BC}_G$. The fixed parameters are listed in the panels.}
\label{bcs_vary}
\end{figure*}

\subsection{Program package {\tt species}}

From the BOSZ theoretical spectra, our synthetic magnitudes in the Vega magnitude system were determined using the corresponding routines of the Python package {\tt species}\footnote{\url{https://species.readthedocs.io/en/latest/index.html}} \citep{2020A&A...635A.182S}. The {\tt species} software package can use the Spanish Virtual Observatory (SVO) filter profile service\footnote{\url{http://svo2.cab.inta-csic.es/theory/fps}}, from which the transmission functions of filters used in instruments of many ground-based observatories and space telescopes can be retrieved. {\tt species} can automatically choose between the AB and Vega photometric systems. The latter choice can be forced by using the spectrum of Vega. In this case {\tt species} gives the magnitude values relative to Vega under the assumption that Vega's brightness is 0.03 mag in all filters by default. While this is mostly true in BVRI filters, it may not be so in other systems. The zero point depends on the absolute flux used to standardize those systems. This is important to point out, because {\tt species} does not take the offsets between systems into account, so the published magnitudes and BCs may have systematic discrepancies with respect to the definition of the particular photometric system in question in the range of a few hundredths of magnitudes in certain filters. At the beginning of the calculation, it downloads the most recent spectrum of Vega from the CALSPEC database \citep{2014PASP..126..711B} and calculates the magnitudes for the synthetic spectra using the magnitudes derived from that. The Vega spectrum used in this paper is {\tt alpha\_lyr\_stis\_011.fits} that can be downloaded from the CALSPEC\footnote{\url{https://ssb.stsci.edu/cdbs/calspec/alpha\_lyr\_stis\_011.fits}} database.

\section{The Scope of the Database}
\label{sec:validating_the_database}

The Spanish Virtual Observatory Filter Profile Service \citep[SVO FPS, ][]{2012ivoa.rept.1015R,2020sea..confE.182R} provides standardized information, including transmission curves and calibration, on more than 7,900 astronomical filters. The service is designed to be compliant to the Virtual Observatory Photometry Data Model, and all the information is provided both as a web portal and VO services so that other services and applications can access the relevant properties of a filter in a simple way.

We carefully checked the SVO FPS and selected the facility/instrument pairs we considered most important in today's astronomical photometry. Our sample contains a total of 752 filters of 78 instruments from 35 ground-based observatories and space telescopes. From the available filter transmission curves we have selected only those that had labels indicating that they can be used for full-fledged photometric measurements. The facilities, instruments and the number of filters are listed in Table~\ref{fac_inst_filt_num}. All filters selected had a minimum wavelength larger than 50 nm and a maximum wavelength smaller than 32 microns in order to fit inside the wavelength range covered by the BOSZ spectra.

The parameter space covered by our BC database is almost the same as that of the BOSZ spectral library. The main difference is that in the spectral library the step size of \teff\ is 100~K below 4000~K, 250~K between 4000 and 12000~K, and 500~K above 12000~K, while in our BC database it is 50~K throughout the temperature range. The magnitude and BC values on the denser grid points were determined by a third-order spline interpolation. Due to potential large changes between the flux of different model spectra, we did not interpolate the spectra themselves to determine the magnitude and BC values for the 50 K step interval but performed the interpolation using the calculated magnitude and BC values at the original grid points. No interpolation was performed for the other parameters (\logg, \am, \cm, $A_V$). The parameters of the final grid are listed in Table \ref{grid_parameters_for_BC_calc}. The Sun bolometric corrections and synthetic magnitudes for all filters can be found in Table \ref{solar_bcs_and_mags}.

\begin{figure*}
\centering
\includegraphics[width=7.2in,angle=0]{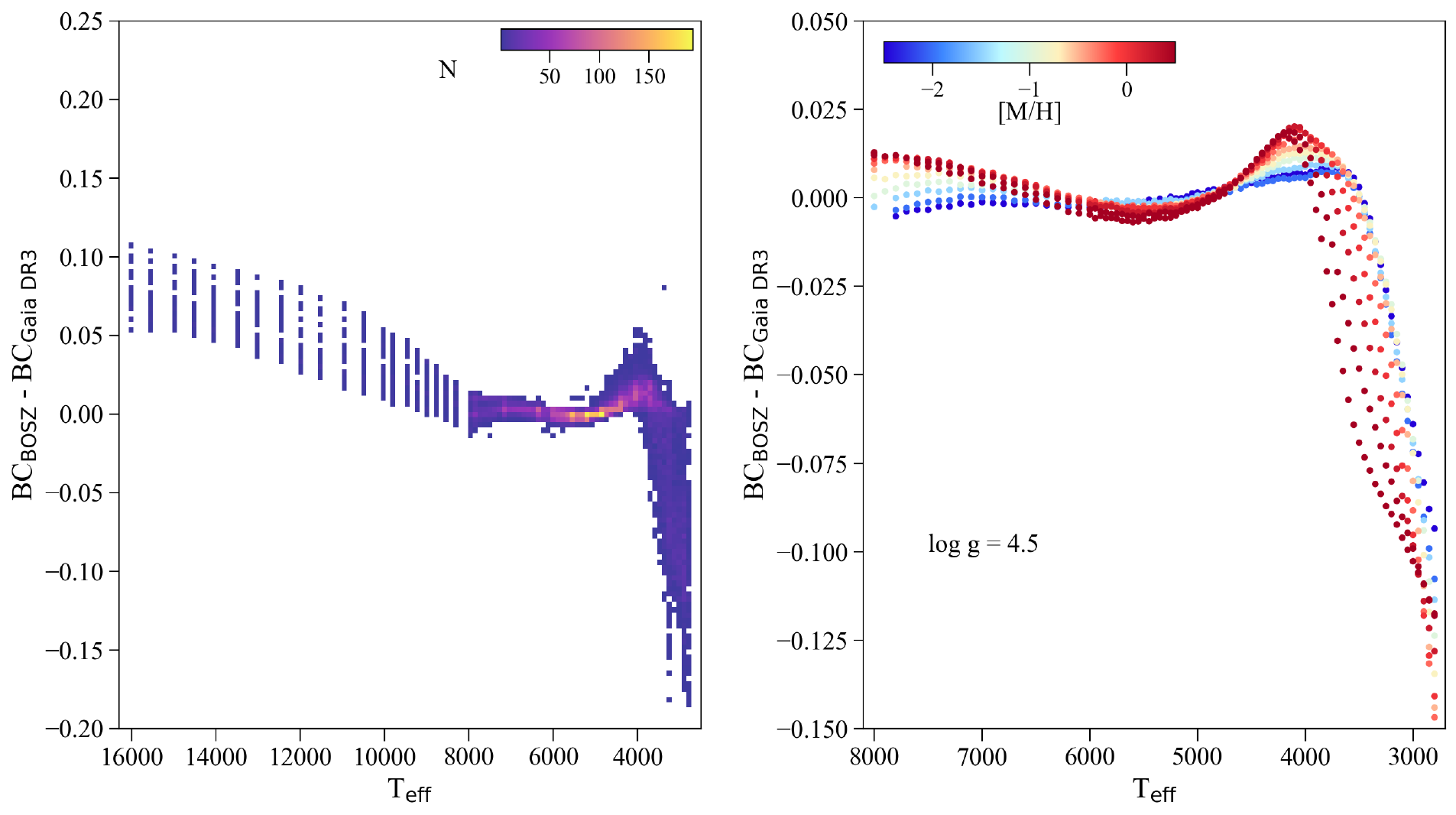}
\caption{Difference between the $\mathrm{BC}_G$ from this paper and from Gaia DR3 as a function of \teff. The left panel shows all commom \teff, \logg, and \mh\ combinations, while in the right panel the \logg\ was set to 4.5 to reveal the metallicity dependent differences on the main sequence. \am\ = \cm\ = 0.0 in both cases.}
\label{bcs_comp}
\end{figure*}

\subsection{Validation of color-color relationships}
\label{validate}

In order to check the accuracy of our synthetic magnitude scale, we compare the observed Gaia--2MASS \citep{2006AJ....131.1163S} color relationship data with the same colors in our database. The observed and modeled Gaia and 2MASS color indices $G_\mathrm{BP} - G_\mathrm{RP}$ and $J - K_\mathrm{s}$ are plotted as a function of all relevant physical parameters (\teff, \logg, \mh, \am, \cm\ and $A_V$) in Fig.~\ref{color2}. We selected nearly two million common stars from the Gaia DR3 and 2MASS databases, which were plotted with contour lines in Fig.~\ref{color2}. To generate the contours, we used only those stars that are brighter than magnitude 12 in the $G$ filter, their error in $G$, $G_\mathrm{BP}$ and $G_\mathrm{RP}$ magnitudes is less than $0.01$, their error in $J$, $H$ and $K_\mathrm{s}$ magnitudes is less than $0.03$, and their error in parallaxes is also less than 10\%. The filtering conditions for errors in brightness values are based on the relationships given by Carrasco\footnote{\url{https://gea.esac.esa.int/archive/documentation/GDR2/Data_processing/chap_cu5pho/sec_cu5pho_calibr/ssec_cu5pho_PhotTransf.html}} for Gaia DR2. As a result of the cuts, 1\,984\,111 Gaia--2MASS stars were included in the sample, from which the contour lines were drawn on the color-color diagrams. We did not filter for interstellar reddening as the synthetic color indices for all reddening values are plotted in the figure.

Overall, the agreement between the synthetic colors and the observed ones is very good, but there are regions in the color space where the synthetic colors do not match the observation. One such region is the high temperature stars where either the $G_\mathrm{BP} - G_\mathrm{RP}$ color is too red, or the $J - K_\mathrm{s}$ color is too blue (or the combination of the two) compared to the color derived from the observations, though the discrepancy is small. The reason behind this discrepancy is currently unknown. The distribution of the synthetic color points forms a "fan" shape much wider than the observation if $G_\mathrm{BP} - G_\mathrm{RP} > 1.5$, where the temperature is lower than 5000$-$7000~K, while the colors have a narrow distribution if $G_\mathrm{BP} - G_\mathrm{RP} < 1.5$ (the temperature panel of Fig.~\ref{color2}). This happens because at larger temperatures than 7000~K the rest of the main atmospheric parameters have a much smaller effect on the stellar spectra than at lower temperatures.

The majority of the observed stars are main sequence stars with solar metallicity (lower branch of the contour lines), and giant stars (upper branch of the contour lines), as can be seen in the \logg\ panel of Fig.~\ref{color2}. Because the Gaia-2MASS sample is restricted in brightness, the sample does not contain stars with $G_\mathrm{BP} - G_\mathrm{RP} > 4$, which are the very low temperature main sequence and giant stars. The data points at the bottom of the fan, shown with blue points in the metallicity panel, represent the metal-poor red dwarfs, which are not in the observed Gaia-2MASS sample.

Some of the discrepancy between synthetic and observed colors come from the fact that the BOSZ database includes main atmospheric parameters where there are no observed stars in order to have a complete grid. One such region is the most metal-rich and $\alpha$-poor combination, which can be seen as a narrow strip of dark blue points in the $\alpha$ panel of Fig.~\ref{color2} between $\mathrm{J}-\mathrm{K}_{\mathrm s}$ 1.5 and 2.2. The reddest, extremely metal-rich (\mh\ = 0.75), but non-existing red giants are at $G_\mathrm{BP} - G_\mathrm{RP} > 4$ and $\mathrm{J}-\mathrm{K}_{\mathrm s} > 1.5$.

\subsection{Effect of grid parameters on the bolometric correction}
\label{parameffect}

Fig.~\ref{bcs_vary} shows the effect of the astrophysical parameters \logg, \mh, \am, \cm, and $A_V$ on Gaia DR3 $G$ bolometric correction ($\mathrm{BC}_G$). In each of the five panels, the values of $\mathrm{BC}_G$ are plotted against the effective temperature, which varies between 6000~K and 2800~K, and the effects of the other four parameters are indicated by color codes. The fixed parameters are shown in the lower left part of the panels in Fig.~\ref{bcs_vary}. For \logg, the abundance values were set to the solar value of 0.0~dex, while for the abundance panels the \logg\ value was set to 2.0~dex, i.e. the figure only shows the case of giant stars. In each case, the bolometric correction starts from around 0.0 at the higher-temperature end of the range, then starts to decrease sharply around 4500~K and becomes increasingly negative. It can be clearly seen that below this temperature, all four atmospheric parameters (\logg, \mh, \am, \cm) have an effect on the bolometric correction, which can even exceed 1.5 magnitudes below 3000~K. In most cases \logg\ and \mh\ have the largest effect on the bolometric correction, but as the temperature decreases \am\ and \cm\ become important factors for cool stars.

Previously the effect of carbon on the Gaia DR3 $G$ bolometric correction has not been studied. By using the BOSZ database, it became possible to examine how \cm\ values affect the bolometric correction. This can be seen in the fourth panel in Fig.~\ref{bcs_vary}. While low \cm\ values have little effect on the bolometric correction, at \cm\ larger than 0$-$0.25~ dex can affect the value of the bolometric correction almost as much as, if not more, than the \am\ dimension. Below 4500~K carbon has a significant effect on the structure of stellar atmospheres that must be taken into account during the model atmosphere calculations, as it was in the BOSZ grid. Thus, for carbon-enriched stars it is important to address their carbon content when calculating their luminosity and radius.

Interstellar reddening has a significant effect on bolometric corrections and should not be ignored, as emphasized, for example, by \cite{2019A&A...632A.105C}. As mentioned above, the spectra themselves were reddened instead of bolometric corrections. The effect of reddening between $A_V$ = 0.0 and 3.1 on the Gaia DR3 $G$ bolometric correction can be seen in the last panel of Fig.~\ref{bcs_vary}. Higher reddening significantly decreases the bolometric correction, although the correlation between the two parameters is not linear.

\begin{figure}
\centering
\includegraphics[width=3.52in,angle=0]{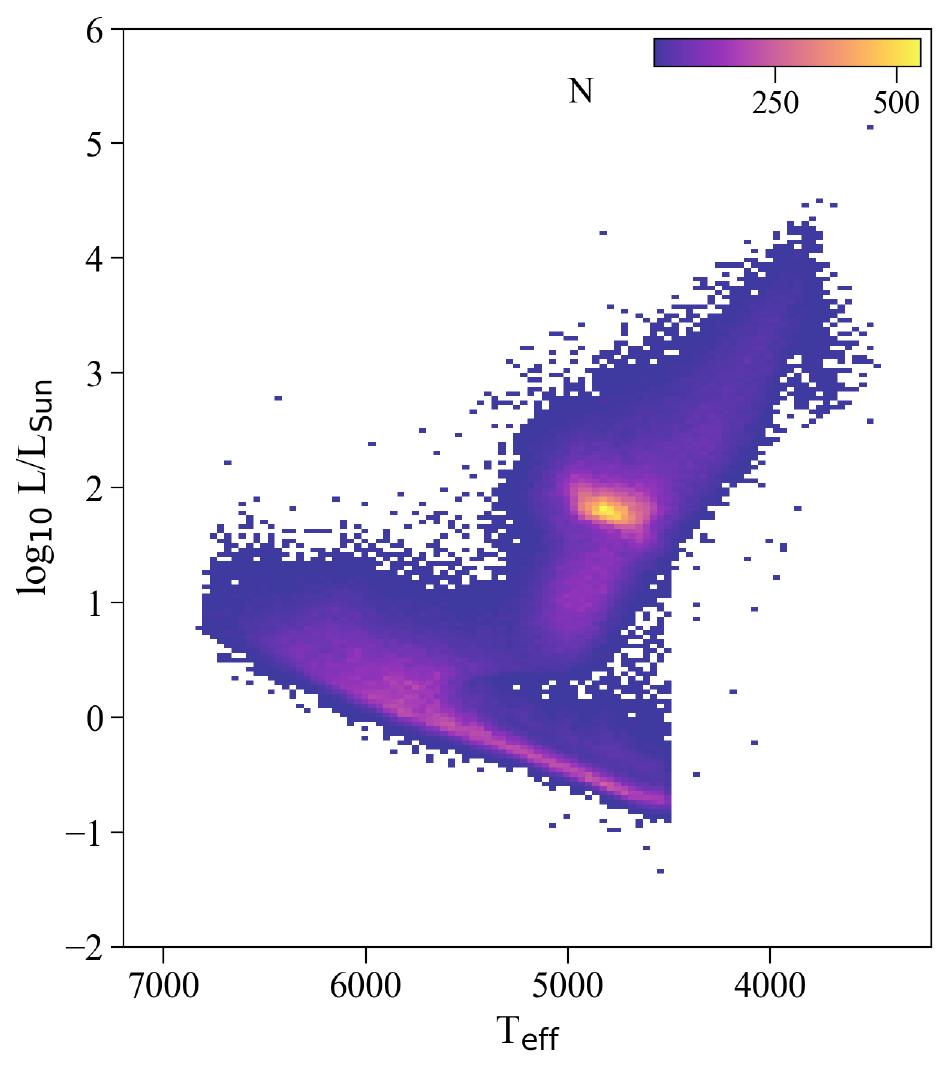}
\caption{The HRD of our selected APOGEE targets using the new Gaia DR3 $\mathrm{BC}_G$ values presented in this paper. }
\label{hrd_final}
\end{figure}

\section{Comparing Gaia DR3 \textit{G} bolometric corrections with literature}
\label{sec:comparision_with_gaia_dr3}

Here, we discuss the differences in the official Gaia DR3 $G$ bolometric corrections \citep{2023A&A...674A..26C} and the ones presented in this paper. The Gaia DR3 bolometric correction software tool also used a MARCS model grid of \citet{2008A&A...486..951G} up to 8000~K (and other spectrum libraries above that value, see Table 2 in \cite{2023A&A...674A..26C}), with a slightly wider range of \teff\ than our new BOSZ models, from 2500 to 20,000 K. The values of \logg\ differ only at the upper limit, ranging up to 5.0 for the Gaia DR3. Its metallicity range of $-5.0$ to $+1.5$ is wider than the range of the new BOSZ models, however, the BOSZ models cover a wider range of values of \am. In \citet{2023A&A...674A..26C}, the $\alpha$ elements enhancement range spans from $0.0$ to $+0.4$. The most significant difference between the two model grids is that the new BOSZ models take into account the contribution of carbon, whereas the Gaia DR3 $G$ bolometric corrections do not.

\begin{figure*}
\centering
\includegraphics[width=7.2in,angle=0]{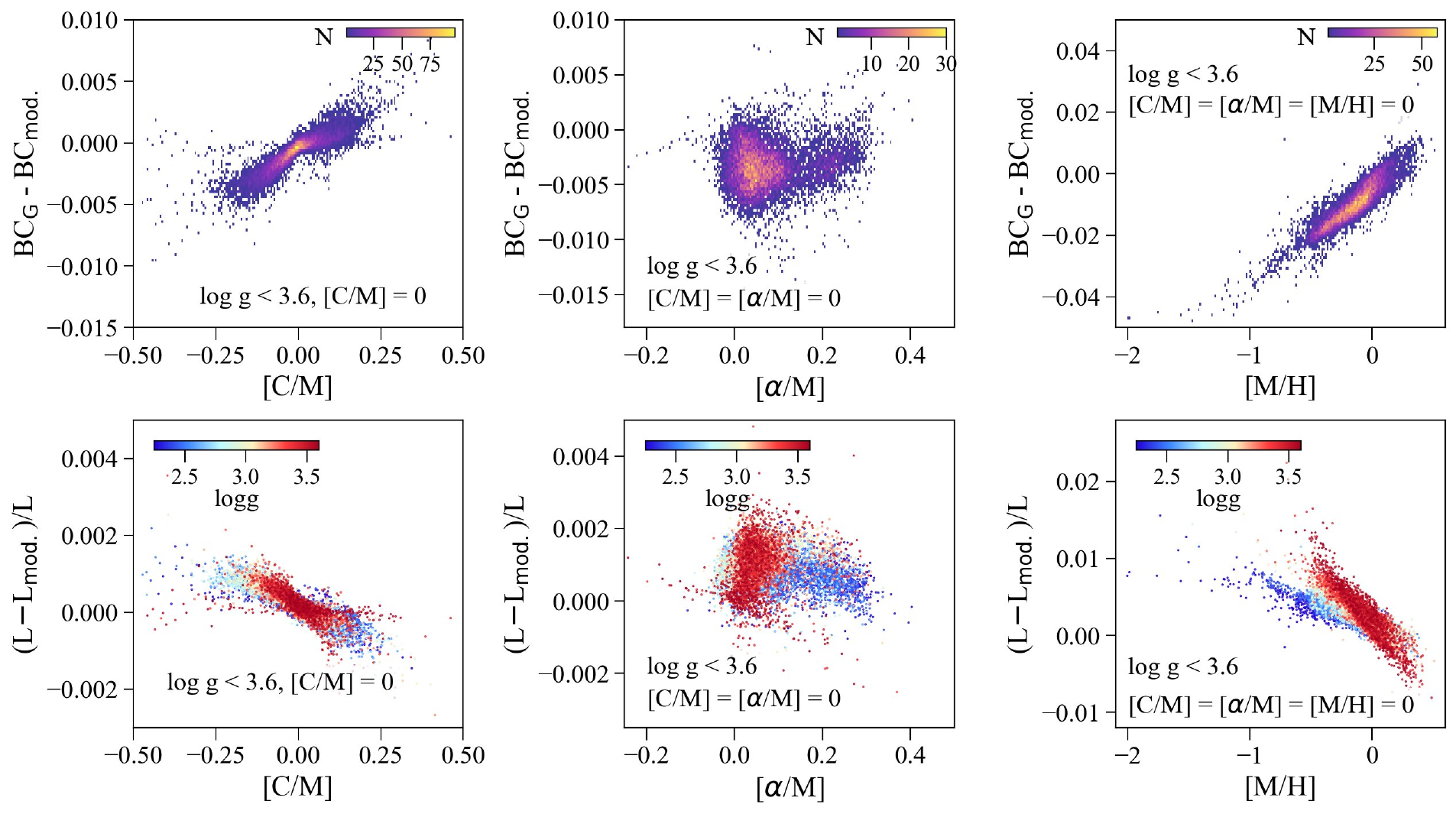}
\caption{The error caused in bolometric correction (top panels) and luminosity (bottom panels) when keeping carbon, alpha, and metallicity at zero for stars with $\log g < 3.6$. The values $\mathrm{BC}_G$ and $L$ were calculated with all three parameters taken into account, and the mod values are when one or more of the parameters \mh, \ \am, \ \cm\ values were fixed to zero. $N$ denotes the number of stars for BC.}
\label{param_effect}
\end{figure*}

The difference between the Gaia DR3 $G$ bolometric corrections and those of \citet{2023A&A...674A..26C} for the same parameters is shown in Fig.~\ref{bcs_comp} as a function of \teff. To explore the effects of updates made on the BOSZ grid compared to the synthetic spectra of \citet{2008A&A...486..951G}, the constant offset of 0.2 applied to Gaia DR3\footnote{See \url{https://gitlab.oca.eu/ordenovic/gaiadr3_bcg/-/blob/main/gdr3bcg/bcg.py}.} was removed. Besides this, their adopted value for the bolometric correction of the Sun is $\mathrm{BC}_{G,\odot} = +0.08$~mag, while ours is directly coming from the model and is equal to 0.0464 mag (see Tab.~\ref{solar_bcs_and_mags}). By correcting for this offset too, the difference seen in Fig.~\ref{bcs_comp} is only due to differences in the synthetic spectral grid used in the calculations.

A large systematic offset that slightly correlates with temperature can be seen for the hot stars above 8000~K. This is probably the result of differing model atmospheres, as Gaia DR3 used the line-by-line opacity stellar model atmospheres from \citet{2004A&A...428..993S}, while the BOSZ database used ATLAS9 models from \citet{2012AJ....144..120M}. The right panel of Fig.~\ref{bcs_comp} shows the \teff\ $<$ 8000~K range in more detail. Although the agreement between the two bolometric corrections is smaller than 0.02 magnitude in the 8000$-$4000~K range, significant temperature dependent systematic offset can be seen again below 4000~K. This is the temperature regime where molecules dominate the spectrum, and the BOSZ grid implemented newer, more up-to-date molecular line lists mostly from the EXOMOL project \citep{2017JQSRT.203..490B, 2017JQSRT.187..453B}, than what was available for \citet{2008A&A...486..951G}.

\section{APOGEE luminosities}
\label{sec:apogee_luminosities}

\subsection{Our APOGEE sample}

In order to determine new luminosity values for the APOGEE survey \citep{2017AJ....154...94M}, we first compiled a list of stars detected by both Gaia DR3 and APOGEE DR17 \citep{2022ApJS..259...35A}, and then filtered this list according to several criteria. Only APOGEE targets without the BAD flag are included in the list. To filter out binaries and pulsating variables, we considered only targets for which the APOGEE parameter $v_\mathrm{scatter}$ (the scatter of radial velocities derived from spectra taken at different times) is less than $1\,\mathrm{km}\,\mathrm{s}^{-1}$. In addition, we also required that the error in individual radial velocities not be greater than $1\,\mathrm{km}\,\mathrm{s}^{-1}$. Finally, we removed targets with a parallax error greater than 10\%. This process left 192\,049 stars in the sample.

After filtering as described above, we applied additional cuts, because $\alpha$ and carbon abundances are unreliable in certain parameter ranges \citep[for more details, see][]{2020AJ....160..120J, 2025AJ....170...96M}. For effective temperatures below 4500~K, only stars with a value of \logg\ less than 3.0 were considered, because carbon abundances of red dwarfs are not reliable. As explored by \citet{2025AJ....170...96M}, stars with effective temperatures lower than 4250~K and \cm\ higher than 0.1 have also unreliable abundances, thus it is necessary to exclude cool, carbon-rich stars from the sample.  After cutting, the sample eventually was left with 177\,728 targets that met the requirements. It should be noted that this sample still contained some metal- and carbon-rich stars above 4250~K, \cm\ values reaching 0.25~dex.

Fig.~\ref{hrd_final} shows the new luminosities as functions of the effective temperature for our APOGEE sample calculated using equations (\ref{L_per_Lsun}) and (\ref{R_per_Rsun}). Here, $\mathrm{BC}_G$ values were derived with radial basis function interpolation using the calibrated astrophysical parameters \teff, \logg, \mh, \am, raw (uncalibrated) \cm\, and \ebv\ published in APOGEE DR17. Because we excluded the red dwarfs from the sample; a vertical straight boundary can be observed at 4500~K, below which only a handful of stars can be seen due to probable incorrect luminosity calculation.

\subsection{The effect of carbon on bolometric correction and  luminosity}

Fig.~\ref{param_effect} shows the effect on the bolometric correction (top) and luminosity (bottom) if we do not take certain parameters into account when determining the BC value. Only RGB stars with \logg\ $ < 3.6$ are selected here to remove erroneous measurements of abundances in the main sequence. On the vertical axis, the values $\mathrm{BC}_G$ and $L$ were calculated with all three parameters taken into account, and the mod values are when one or more of the parameters \mh, \am, \cm\ values were fixed to zero. As indicated in the figure, in the first column \cm\ was set to 0.0, but \mh, and \am\ were taken from APOGEE, in the second column \cm\ = \am\ = 0.0, but metallicity was allowed to vary, in the last column all three parameters were zeros, which means only \teff, \logg, and the reddening was taken into account.

The effect on BCs is given as a magnitude difference, whereas the effect on luminosities is expressed in relative terms. The histograms are colored by the number of stars. It can be clearly seen that ignoring all three parameters can cause errors of up to 0.04 magnitude and up to 1\% in the bolometric correction and luminosity over the entire temperature range considered. It should be noted that there is a clear correlation between the bolometric correction difference and the \mh\ value of the star. Of course, the errors decrease once the metallicity and $\alpha$ elements are considered, but if carbon is not used in the BC calculation, at 4500--5000~K, where most of the stars are, there still remains around 0.005 mag and 0.2\% error in BC and luminosity, respectively. It can be clearly seen that there is a correlation between BC differences and the \cm\ value when \am\ and \mh\ are considered, but \cm\ is not (top left panel of Fig.~\ref{param_effect}). These trends are also mirrored in the case of luminosity, with the opposite sign. This is an expected behavior, because as we have shown in Section~\ref{parameffect} the value of $\mathrm{BC}_G$ increases with increasing \cm. Thus, ignoring carbon will result in lower $\mathrm{BC}_G$ values than including it, decreasing the luminosity.

\section{Overview}
\label{sec:overview}

In this paper, we present a new, extensive database of synthetic stellar magnitudes and bolometric corrections based on the latest BOSZ spectral library. Our goal is to improve estimates of stellar luminosities by delivering accurate BCs across hundreds of photometric filters. In the Gothard Observatory Synthetic Stellar Photometry Database synthetic magnitudes were computed in the Vega  system using the Python package {\tt species}, and BCs were derived from fundamental flux--magnitude relationships, by applying the interstellar reddening directly to the spectra via Cardelli’s extinction law. Covering 752 filters from 78 instruments (ground- and space-based), the database spans effective temperatures from 2800 to 16000~K, \logg\ from $-0.5$ to 5.5, metallicity from $-2.5$ to 0.75, \am\ from $-0.25$ to 0.5, \cm\ from $-0.75$ to 0.5, and reddening up to A$_{\rm V}$ = 3.1 mag. Using high-resolution (R = 50000) synthetic spectra allowed us to precisely track the effect of abundances on the bolometric corrections and luminosity of stars.

Validation against nearly two million Gaia–2MASS stars shows strong agreement in color–color space, with minor discrepancies at extreme temperatures. Compared to existing Gaia DR3 BCs, our corrections differ by up to 0.2 mag for hot stars, and up to 0.12 mag below 4000~K probably due to improved treatment of molecular opacities in the BOSZ database. By applying the new BCs to 192\,000 APOGEE stars, we calculated refined luminosities, and also demonstrated that neglecting carbon or $\alpha$ enhancements can introduce up to 1 \% errors in luminosity. Overall, the new Gothard Observatory Synthetic Stellar Photometry Database may enable more accurate fundamental parameter determinations for large stellar samples using a vast amount of past, present, and upcoming surveys, such as Gaia, LSST, and the Roman Space Telescope.

\begin{acknowledgements}
This project has been supported by the LP2021-9 Lend\"ulet grant of the Hungarian Academy of Sciences. On behalf of the "Calculating the Synthetic Stellar Spectrum Database of the James Webb Space Telescope" project, we are grateful for the possibility to use HUN-REN Cloud (see \citep{heder2022}; \url{https://science-cloud.hu}) which helped us achieve the results published in this paper. This project has received funding from the HUN-REN Hungarian Research Network.

Funding for the Sloan Digital Sky Survey IV has been provided by the Alfred P. Sloan Foundation, the U.S. Department of Energy Office of Science, and the Participating Institutions.

SDSS-IV acknowledges support and resources from the Center for High Performance Computing at the University of Utah. The SDSS website is www.sdss.org.

SDSS-IV is managed by the Astrophysical Research Consortium for the Participating Institutions of the SDSS Collaboration including the Brazilian Participation Group, the Carnegie Institution for Science, Carnegie Mellon University, Center for Astrophysics | Harvard \& Smithsonian, the Chilean Participation Group, the French Participation Group, Instituto de Astrof\'isica de Canarias, The Johns Hopkins University, Kavli Institute for the Physics and Mathematics of the Universe (IPMU) / University of Tokyo, the Korean Participation Group, Lawrence Berkeley National Laboratory, Leibniz Institut f\"ur Astrophysik Potsdam (AIP),  Max-Planck-Institut f\"ur Astronomie (MPIA Heidelberg), Max-Planck-Institut f\"ur Astrophysik (MPA Garching), Max-Planck-Institut f\"ur Extraterrestrische Physik (MPE), National Astronomical Observatories of China, New Mexico State University, New York University, University of Notre Dame, Observat\'ario Nacional / MCTI, The Ohio State University, Pennsylvania State University, Shanghai Astronomical Observatory, United Kingdom Participation Group, Universidad Nacional Aut\'onoma de M\'exico, University of Arizona, University of Colorado Boulder, University of Oxford, University of Portsmouth, University of Utah, University of Virginia, University of Washington, University of Wisconsin, Vanderbilt University, and Yale University.

This research has made use of the SVO Filter Profile Service "Carlos Rodrigo", funded by MCIN/AEI/10.13039/501100011033/ through grant PID2020-112949GB-I00

\end{acknowledgements}

\bibliographystyle{aa}
\bibliography{references}

\end{document}